\documentstyle[epsfig,osa,manuscript]{revtex}

\begin{document}

\title{Probability Density Functions of Decaying Passive Scalars 
in Periodic Domains : An Application of Sinai-Yakhot Theory}
\author{Jai Sukhatme \footnote{{\it email : jai@ucar.edu}}}
\address{National Center for Atmospheric Research, Boulder, CO}
\date{\today}
\maketitle

\begin{abstract} Employing the formalism introduced by Sinai and Yakhot 
[PRL, 63(18), p. 1962, 1989], we study the probability density
functions (pdf's) of decaying passive scalars in periodic domains under the influence of 
smooth large scale velocity fields. The
particular regime we focus on is one where the normalized scalar pdf's attain a self-similar
profile in finite time, i.e., the so called strange or statistical eigenmode regime. 
In accordance with the work of Sinai and Yakhot, the central regions of the pdf's are power laws. But
the details of the pdf profiles 
are 
dependent on the physical parameters in the problem. 
Interestingly, for small Peclet numbers the pdf's {\it resemble} stretched or pure 
exponential functions, whereas in the limit of large
Peclet numbers,
there emerges a universal 
Gaussian form for the pdf. 
Numerical simulations are used to verify these predictions.
\end{abstract}

\narrowtext
\section{Introduction} 

We examine the probability density functions (pdfs) of
decaying passive scalars without mean gradients under the action of smooth, incompressible and
time aperiodic flows in bounded periodic domains.  In this situation the evolution of a passive
scalar, $\phi(x,y,t)$, is governed by the advection-diffusion (AD) equation.

\begin{equation} 
\frac{\partial \phi}{\partial t} + (\vec{u} \cdot \nabla) \phi = \kappa \nabla^2
\phi \label{1a} 
\end{equation} 
Here $\kappa$ represents the molecular diffusivity of the passive
scalar and $\vec{u}(x,y,t)$ is the advecting velocity field. The
domain ({\sf D}) under consideration is periodic, specifically we take {\sf D} to be $[0,2\pi]
\times [0,2\pi]$ with opposite sides identified (i.e. a 2-torus). \\

In this work, we consider time aperiodic velocity fields whose spatial scale of variation 
is comparable to the
scale of the domain. Essentially, the flows are of the type encountered in chaotic 
advection \cite{Ottino} and the parameters correspond to relatively large Peclet numbers.
We focus on the case when the normalized scalar pdf's attain a
self-similar profile. This self-similar regime was first described by Pierrehumbert \cite{Ray-94},\cite{Ray-Chaos}
(who named it a strange or statistical eigenmode) and has 
recently been examined in detail by Fereday and Haynes
\cite{FH2} (hereafter FH) and also by Sukhatme and Pierrehumbert \cite{me} (hereafter SP).  The
self-similarity of the normalized pdf is indicative of a non-equilibrium steady state. 
Physically, as was
explained in SP and FH, the ingredients in the balance responsible for this state are : (i) a limit
on how thin filaments can get (as $\kappa > 0$) and (ii) the "folding and filling" of filaments
induced by the finite domain. \\

To place the things in a proper perspective, we introduce the following scales : L - the scale of
the domain, $l_{v}$ - the scale of variation of the velocity field and $l_{s}(t)$ - the
maximum scale of variation of the scalar field.  In terms of these scales the self-similar
strange eigenmode is characterized by $l_{s}(t) \sim l_{v} \sim \L$.  Due to this similarity
of scales the problem possesses a global nature (see FH and SP). Hence,
approximations based on scale separation (such as shifting to a comoving reference frame),
which have yielded excellent results in other smooth advection diffusion regimes - sometimes
referred to as the Batchelor regime -
cannot be fruitfully utilized (see \cite{Chertkov-95}, \cite{BF-99} or \cite{Falk} for a 
recent review). Also, when $l_{s}(t) << l_{v} \sim \L$ (i.e., the Batchelor regime), it has 
been demonstrated that the pdf's are 
non-universal, i.e. their shape
evolves in time \cite{BF-99}. 
Obviously, in such a situation there is no limiting scalar pdf and the theory 
of Sinai and Yakhot \cite{Sinai}, which apriori assumes the existence of such a limit, fails.
\\ 

Indeed, it is the attainment of a self-similar, i.e. limiting, pdf profile in 
finite time that makes the 
strange eigenmode regime a suitable candidate for applying the Sinai-Yakhot 
formalism. \\

\section{The PDF Equation}
Consider the dimensionless normalized variable $X = {\phi}/{<\phi^2>^{1/2}}$, by assuming
the stationarity in time of $<X^{2n}>$ for all $n$, Sinai and Yakhot \cite{Sinai}
showed,

\begin{equation}
(2n-1)< X^{2n-2} ~ \frac{ {(\nabla \phi)}^{2} }{Q_1} > = < X^{2n} >
\label{1c}
\end{equation}
where $Q_1 = < {(\nabla \phi)}^{2} >$. Further utilizing Eq. (\ref{1c}), they showed that
the pdf of $X$ is given by
(denoting the sample space variable by the same symbol),

\begin{equation}
P(X) = \frac{C_1}{g(X)}~\textrm{exp}[ -{\int_0}^{X} \frac{u}{g(u)}du ]
\label{1d}
\end{equation}
where $g(X)$ 
represents the conditional expectation of the normalized dissipation, i.e 
$g(X) = < (\nabla \phi)^2 / Q_1  | X >$. As it turns out, later work 
\cite{Ch1},\cite{Ch2},
\cite{Pope-Ching}, \cite{Dopazo} (see 
\cite{Pope} for an overview) clarified that
the pdf of any statistically homogenous twice differentiable random field, say $\psi(\vec{x},t)$,
is given by \cite{Dopazo}, \cite{Ching-Kraichnan},

\begin{equation}
P(\psi,t) = \frac{C_2}{g(\psi,t)/Q_2}~\textrm{exp}[ \int_{0}^{\psi} \frac{r(u,t)}{g(u,t)}du ]
\label{1e}
\end{equation}
$Q_2=<(\nabla X)^2>$ and $r(\psi,t) = < (\nabla^2 \psi) | \psi>$,  $g(\psi,t) = < {(\nabla \psi)}^2 | \psi >$ 
represent
the conditional diffusion and conditional dissipation respectively. Furthermore, it was
shown that if the moments of $\psi(\vec{x},t)$ are stationary then \cite{Ching-Kraichnan},\cite{Vallino}
\footnote{Ching and Kraichnan \cite{Ching-Kraichnan} hint at the possibility
of attaining stationary normalized momemts by utilizing a cyclic domain, the self-similar
strange eigenmode appears to be precisely this case.\\ },

\begin{equation}
r(\psi,t) = -\frac{< {(\nabla \psi)}^2 >}{< \psi^2 >} \psi 
\label{1f}
\end{equation}
Ofcourse, Eq. (\ref{1f}) when substituted in Eq. (\ref{1e}) yields a pdf similar 
to the Sinai-Yakhot expression,
i.e., Eq. (\ref{1d}). \\

\subsection{Conditional Statistics in the Strange Eigenmode Regime}
In the 
strange eigenmode regime, starting with $l_{s}(t=0) \sim l_{v} \sim L$, 
\footnote{Other initial conditions, especially $l_{s}(t=0) << l_{v} \sim L$ entail
an evolution of the scalar field through distinct regimes (see SP and FH for details).\\}
after a transient period, it is seen that (see SP and FH),

\begin{equation}
< |\phi(x,y,t)|^n > \sim e^{-{\alpha}_n t} ~;~ t> T
\label{1g}
\end{equation}
where $T$ represents the duration of the transient period.
More importantly, $\alpha_n = n\alpha_1$, this linearity 
implies the stationarity of the moments of the normalized scalar field 
\footnote{The exponential decay of moments is also valid when $l_{s}(t) <<
l_{v}$, but in this case $\alpha_n$ is a nonlinear function of the moment order $n$, i.e.
the moments are not stationary \cite{BF-99}.\\}.
Ofcourse, given the stationarity, we are justified in using Eq. (\ref{1f}), with 
$X$ replacing $\psi$. 
Substituting in Eq. (\ref{1f}) from Eq. (\ref{1g}) we get, 

\begin{equation}
r(X,t) = r(X) = -\frac{< {(\nabla X)}^2 >}{< X^2 >} X = -\frac{\alpha_2}{2\kappa} X 
\label{1h}
\end{equation}
Regarding the conditional dissipation, if $X$ and $\nabla X$ are independent, then
$g(X,t) = < (\nabla X)^2 >$. This coupled with the fact that $g(X,t)$ is even led
Sinai and Yakhot to propose 
the expansion (in the vicinity of $X=0$)\cite{Sinai},

\begin{equation}
g(X,t) = g(X) = < {(\nabla X)}^2 > + \beta X^2 + ... ~;~ \beta = \frac{1}{2} \frac{\partial^2 g}{\partial X^2}|_{X=0}
\label{1i}
\end{equation}
Substituting from Eq. (\ref{1g}), we have (to order $X^2$),

\begin{equation}
g(X) = \frac{\alpha_2}{2\kappa} ( 1 + \frac{2\kappa\beta}{\alpha_2} X^2 )
\label{1j}
\end{equation}
Furthermore, the normalized conditional diffusion and dissipation are,
$R(X) = r(X)/< {(\nabla X)}^2 >$ and $G(X) = g(X)/< {(\nabla X)}^2 >$ respectively. Using
Eq. (\ref{1h}) and Eq. (\ref{1j}), 

\begin{equation}
R(X) = -X ~~;~~ G(X) = 1 + \frac{2\kappa\beta}{\alpha_2} X^2
\label{1ll}
\end{equation}
Recent numerical work (see FH) suggests that, $\alpha_2$ tends to a non-zero limit as $\kappa \rightarrow 0$, 
hence Eq. (\ref{1ll}) gives
$G(X) \rightarrow 1 $ assuming that $\beta$ does not overwhelm the
limit. In other words, keeping the assumption regarding $\beta$ in mind, $X$ and $\nabla X$
tend to become independent as $\kappa \rightarrow 0$ and we expect the core of the pdf to tend to a universal
Gaussian profile. \\

\begin{figure} 
\begin{center} 
\epsfxsize=12.0 cm
\epsfysize=12.0 cm
\leavevmode\epsfbox{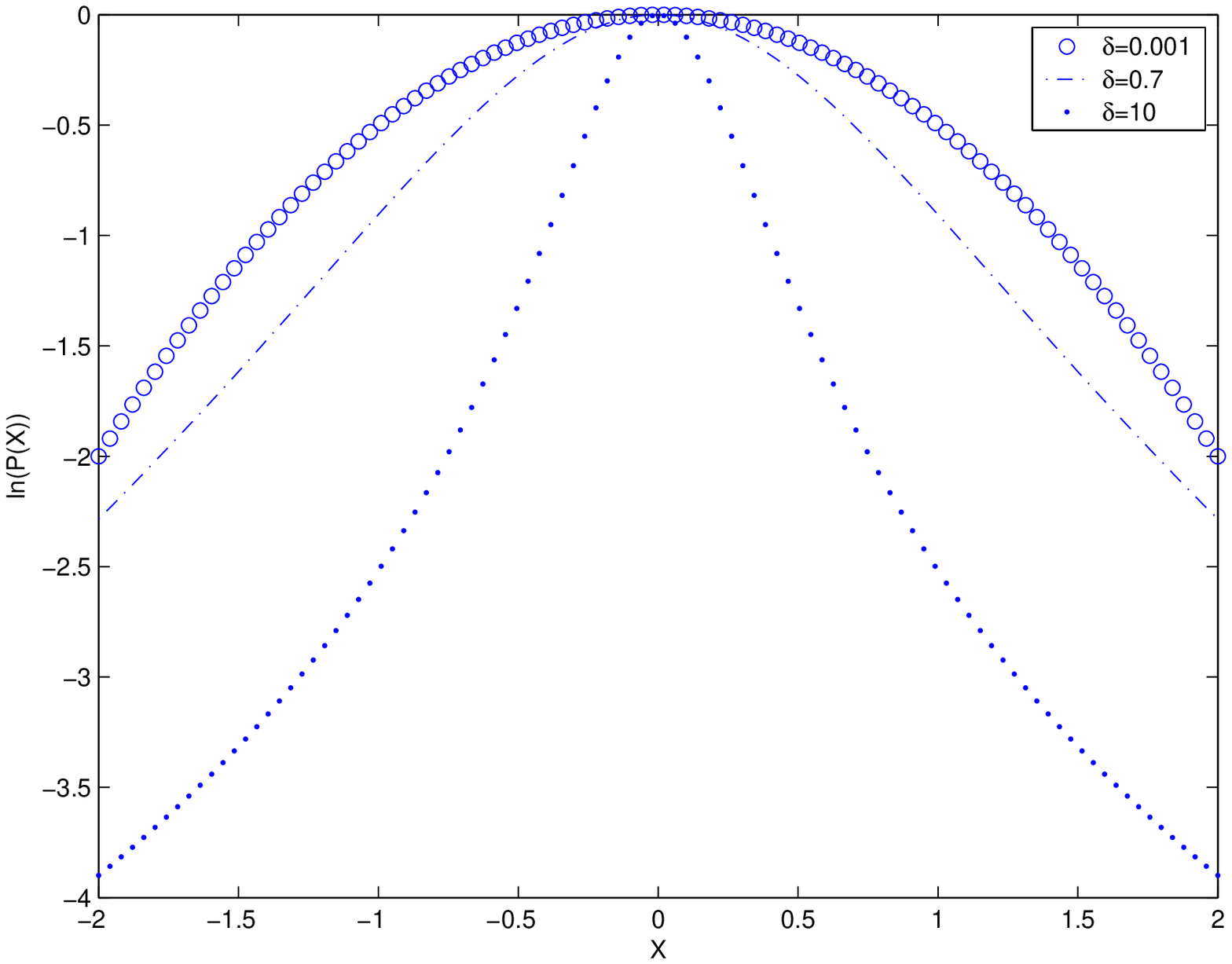}
\end{center}
\caption{$\ln(P(X))$ Vs. $X$ from Eq. (\ref{1l}) for differing $\delta$.} 
\label{fig:fig0} 
\end{figure}

Further substituting 
Eq. (\ref{1h}) and Eq. (\ref{1j}) in Eq. (\ref{1e}) yields 
(the central part of) the pdf of $X$ to be,

\begin{equation}
%P(X) =  \frac{2\kappa C_2}{\alpha_2} [ 1 + \frac{2\kappa\beta}{\alpha_2} X^2 ] ^{-\gamma} ~;~ 
P(X) = C_2~  [ 1 + \frac{2\kappa\beta}{\alpha_2} X^2 ] ^{-\gamma} ~;~ 
\gamma = 1 + \frac{\alpha_2}{4\kappa\beta}
\label{1l}
\end{equation}
Note that, even though the power-law is in agreement with the work of FH, their arguments
apply to the tail of the pdf whereas the above expression is valid in the vicinity of $X=0$. 
For further elucidation, defining $\delta = 2\kappa\beta/\alpha_2$, let us examine how 
the shape of $P(X)$ behaves with $\delta$. In terms of $\delta$,

\begin{equation}
P(X) = C_2~  [ 1 + \delta X^2 ] ^{-\gamma} ~;~ \gamma = 1 + \frac{1}{2\delta}
\label{1mm}
\end{equation}
For large $\delta$ we have $\gamma \rightarrow 1$, $\ln(P(X)) \rightarrow -\ln(1+\delta X^2)$.
As both $\delta$ and $X$ are $O(1)$ quantities, all powers of $X$ contribute to $\ln(P(X))$. 
On the other hand, for small $\delta$ 
we have $\gamma \rightarrow 1/2\delta$ and $\ln(P(X)) \sim -X^2/2$, which is the 
expected outcome
from the earlier discussion.
Profiles for $\delta=10, 0.7, 0.001$ are shown 
in Fig. \ref{fig:fig0}. Note that, as $\delta$ decreases $P(X)$ 
goes from {\it resembling} a
stretched exponential $\rightarrow$ pure exponential $\rightarrow$ (expected) Gaussian function. \\

\section{Numerical Investigation}
The AD equation was approximated
by a lattice map \cite{Ray-Chaos} followed by diffusion in Fourier space.
The
velocity field is of a single large scale, specifically, we employ the sine flow \cite{Alva},\cite{Ant},

\begin{eqnarray}
u(x,y,t) = f(t)~A_1 ~ \textrm{sin}(y + p_n) \quad  \nonumber \\
v(x,y,t) = (1-f(t))~A_2 ~ \textrm{sin}(x + q_n)  \nonumber \\
\label{1m}
\end{eqnarray}
where $f(t)$ is 1 for $nT \le t < (n+1)T/2$ and 0 for $(n+1)T/2 \le t < (n+1)T$. 
$p_n , q_n$ ($\in [0,2\pi]$) are random numbers selected
at the beginning of each iteration, i.e., for each period $T$. $A_1,A_2$ control the strength of the flow.
The flow is implemented as a 
2D lattice map $(x_n,y_n) \rightarrow (x_{n+1},y_{n+1})$ \cite{Ant},\cite{Ray-Chaos}.
A key feature is that the randomness due to $p_n,q_n$ breaks any barriers which 
may, and generically do, exist in 2D area preserving mappings \cite{Meiss}. \\

\begin{figure} 
\begin{center} 
\epsfxsize=12.0 cm
\epsfysize=14.0 cm
\leavevmode\epsfbox{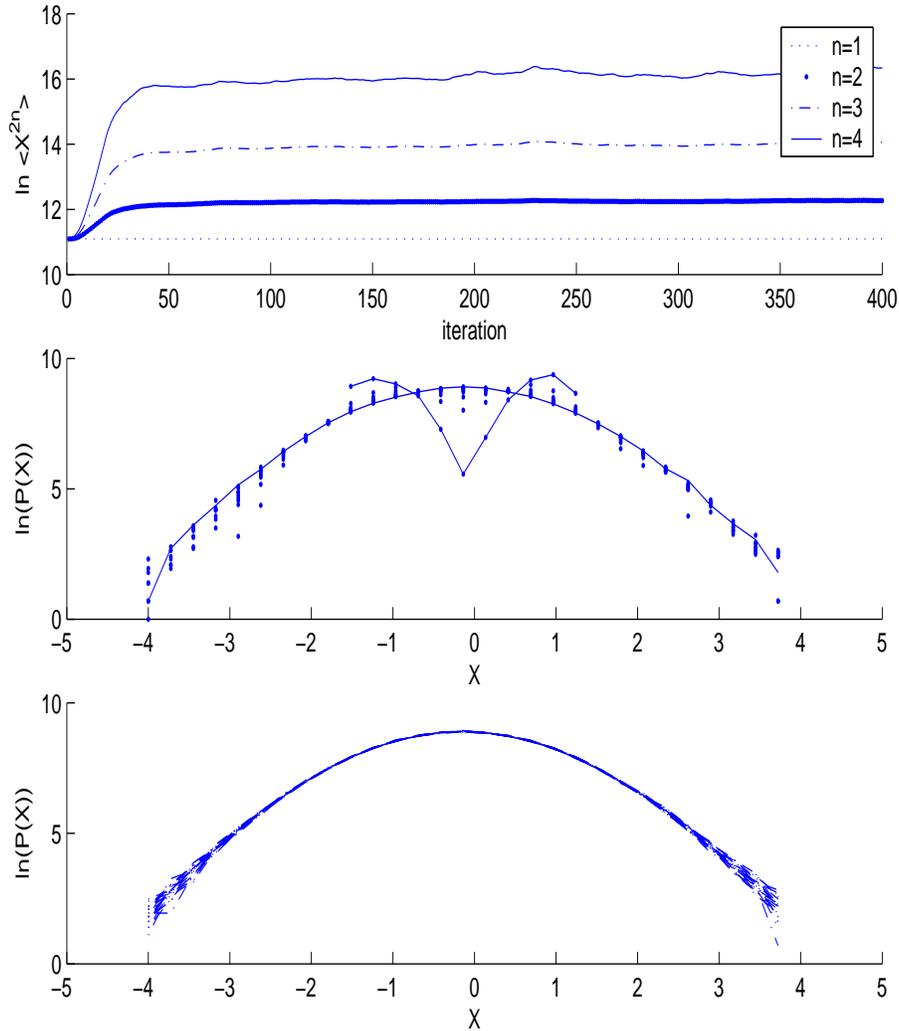}
\end{center}
\caption{Typical evolution scenario with $\kappa=9.33 \times 10^{-4}$. 
Upper panel shows $\ln(<X^{2n}>)$ Vs. iteration for n=1,2,3 and 4. 
The middle panel
shows the pdf's from iteration 10 to 110 (solid lines at iteration 10 and 110)
The
lowermost panel shows the pdf's from iteration 150 to 350, i.e. when the normalized moments have 
become stationary.} 
\label{fig:fig1} 
\end{figure}

{\it A typical example :} Starting with a mean zero checkerboard initial condition 
on a $256 \times 256$ grid ($A_1=A_2=2, \kappa=9.33 \times 10^{-4}$), 
a typical evolution 
scenario is shown in Fig. \ref{fig:fig1}. As is seen, the normalized moments attain
constant values after a transient period of about 70 iterations 
\footnote{The transient period may appear large, but it is important to note that 
the strange eigenmode appears only after the scalar field has folded and filled the domain (see SP and FH 
for details). Whereas, 
significant decay of the variance starts much before this time, specifically when the diffusive scale is 
reached.\\}. During this transient period $P(X)$
evolves (a remnant of the initial double delta pdf can be seen at iteration 10). Finally,
after the moments become stationary, $P(X)$ attains a self-similar profile as is seen in lowermost panel
of Fig. \ref{fig:fig1}. 
\footnote{It is important to keep in mind that these results are for large Peclet numbers. In fact, 
in numerical runs with smaller Peclet numbers, 
even at large times, $<X^{2n}>$ fluctuates with a fairly large amplitude . \\} \\

{\it Peclet number dependence :} To investigate the effect of changing the Peclet number, we run 
a set of simulations with : (a) fixed flow strength and varying diffusivity and (b) fixed 
diffusivity and varying flow strengths. \\

\begin{figure} 
\begin{center} 
\epsfxsize=12.0 cm
\epsfysize=12.0 cm
\leavevmode\epsfbox{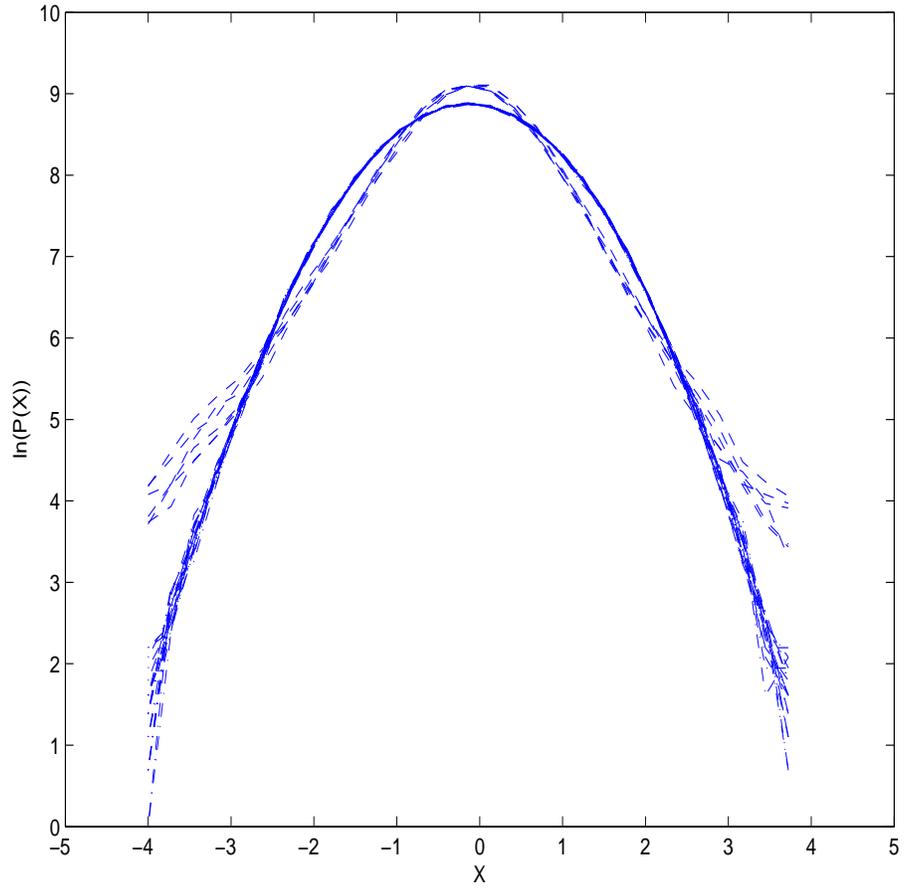}
\end{center}
\caption{Self-similar pdf's for $\kappa=2 \times 10^{-3}$ (dashed), $\kappa=5.78 \times 10^{-4}$ (solid)} 
\label{fig:fig2} 
\end{figure}

\begin{figure}
\begin{center}
\epsfxsize=12.0 cm
\epsfysize=12.0 cm
\leavevmode\epsfbox{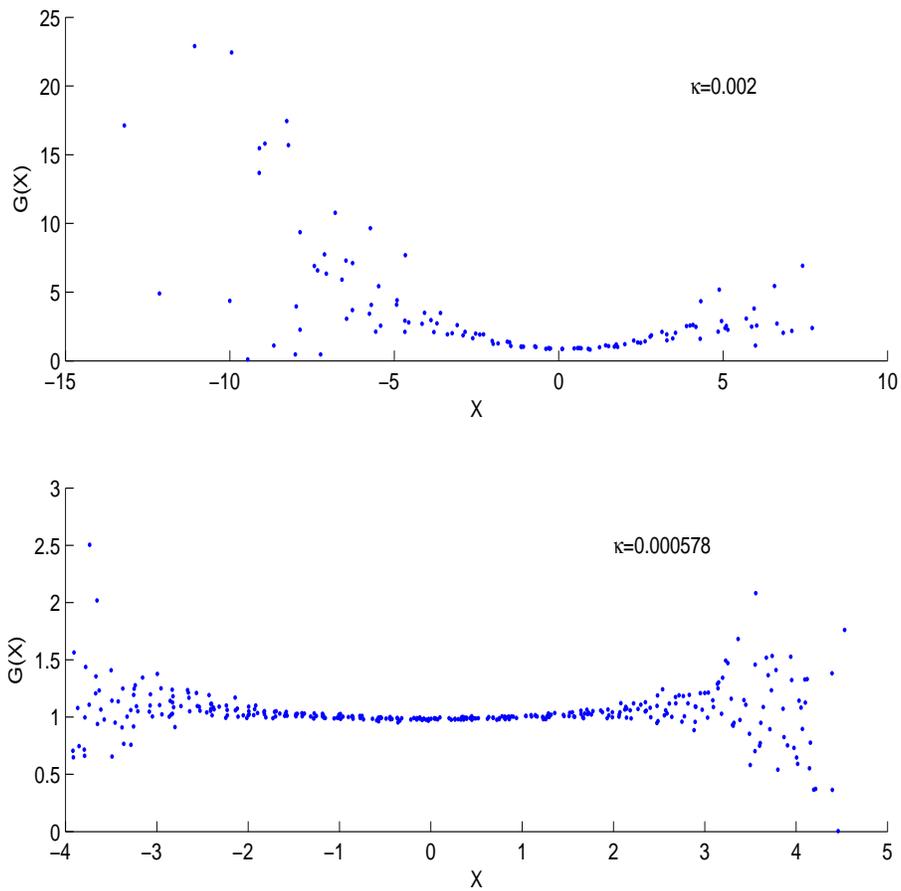}
\end{center}
\caption{The normalized conditional dissipation ($G(X)$) Vs. $X$ for the same set of diffusivities as
in Fig. \ref{fig:fig2}.
Note the range of the axes in the two subplots.}
\label{fig:fig3}
\end{figure}

\begin{figure} 
\begin{center} 
\epsfxsize=12.0 cm
\epsfysize=12.0 cm
\leavevmode\epsfbox{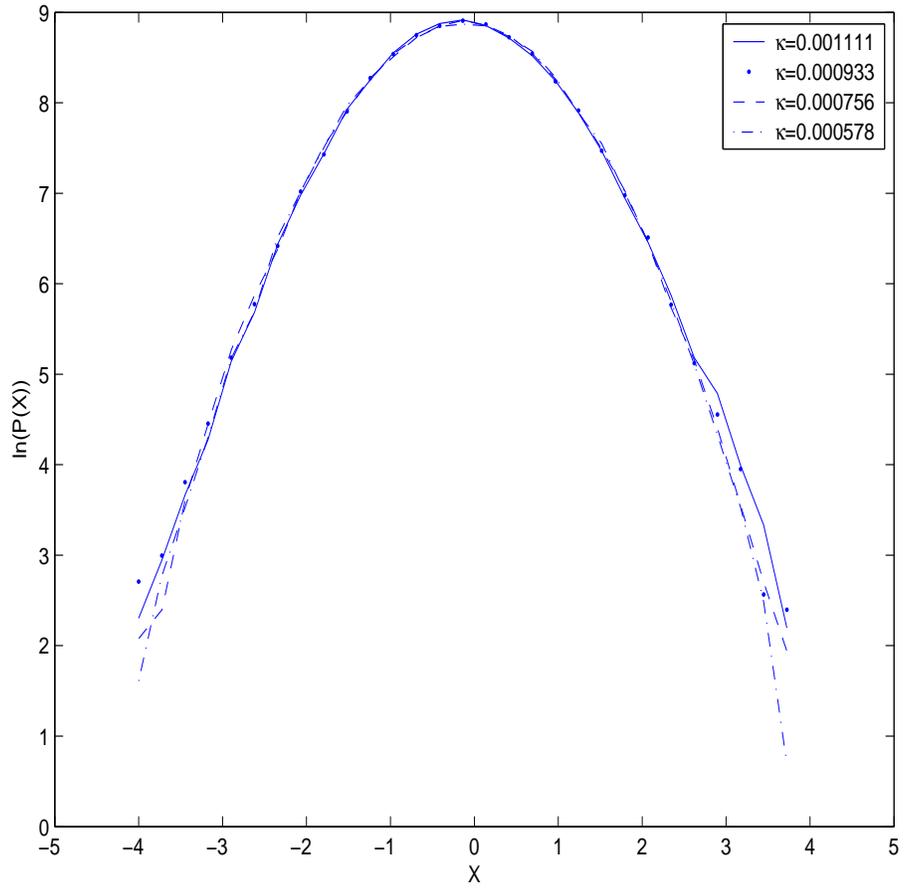}
\end{center}
\caption{pdfs for $\kappa=1.11 \times 10^{-3}, 9.33 \times 10^{-4}, 7.56 \times 10^{-4}, 5.78 \times 10^{-4}$ :
The emergence of a universal Gaussian core.}
\label{fig:fig4} 
\end{figure}

\begin{figure} 
\begin{center} 
\epsfxsize=12.0 cm
\epsfysize=12.0 cm
\leavevmode\epsfbox{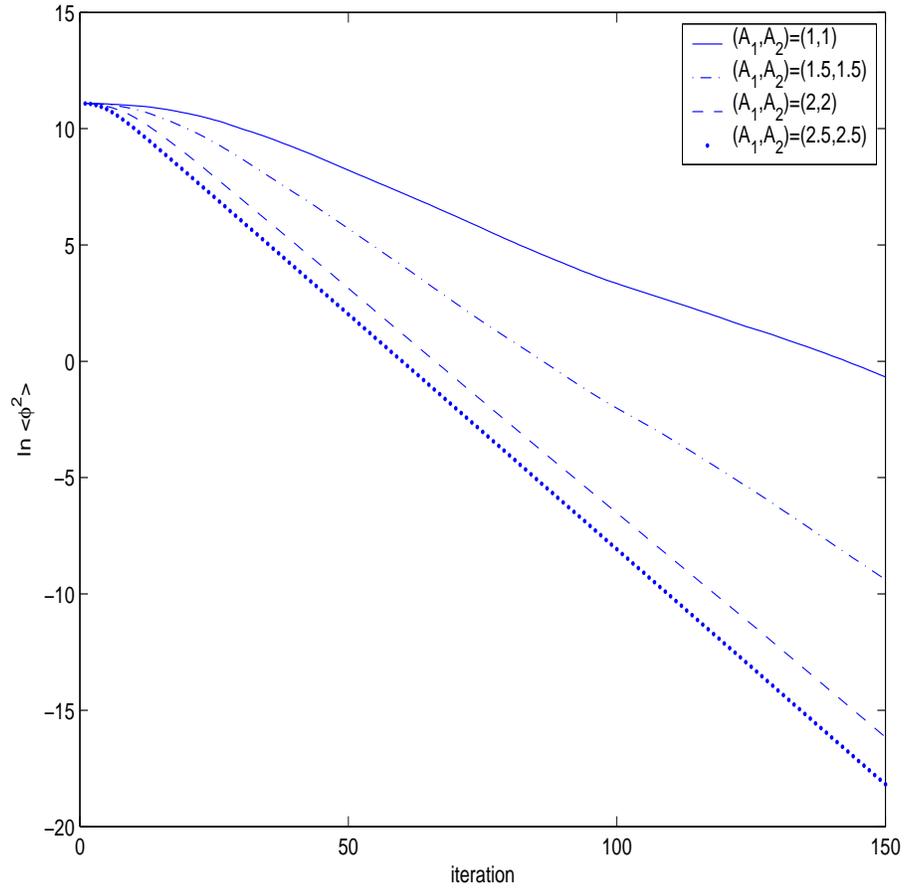}
\end{center}
\caption{Decay of the variance with fixed $\kappa$ and varying flow strengths. 
Clearly, $\alpha_2 \propto$ (flow strength).}
\label{fig:fig5} 
\end{figure}

\begin{figure} 
\begin{center} 
\epsfxsize=12.0 cm
\epsfysize=12.0 cm
\leavevmode\epsfbox{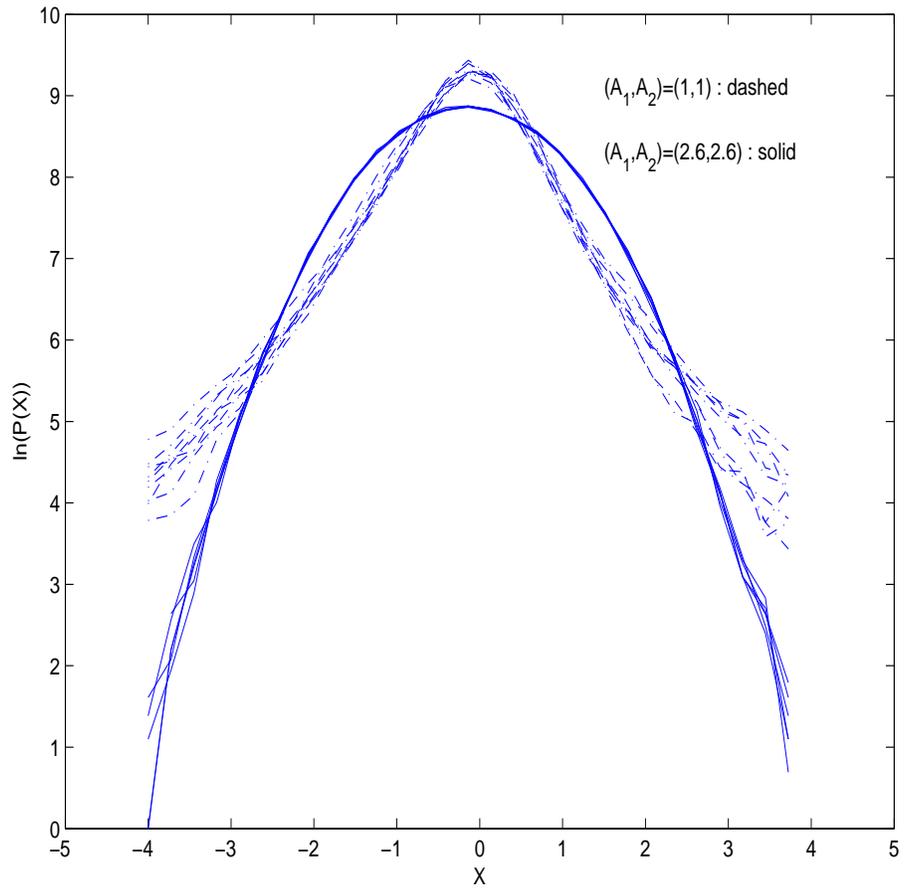}
\end{center}
\caption{$\ln(P(X))$ Vs. $X$ in the self-similar stage for $(A_1,A_2)=(1,1)$ (dashed curves) and $(A_1,A_2)=(2.6,2.6)$.}
\label{fig:fig6} 
\end{figure}

\begin{figure} 
\begin{center} 
\epsfxsize=12.0 cm
\epsfysize=12.0 cm
\leavevmode\epsfbox{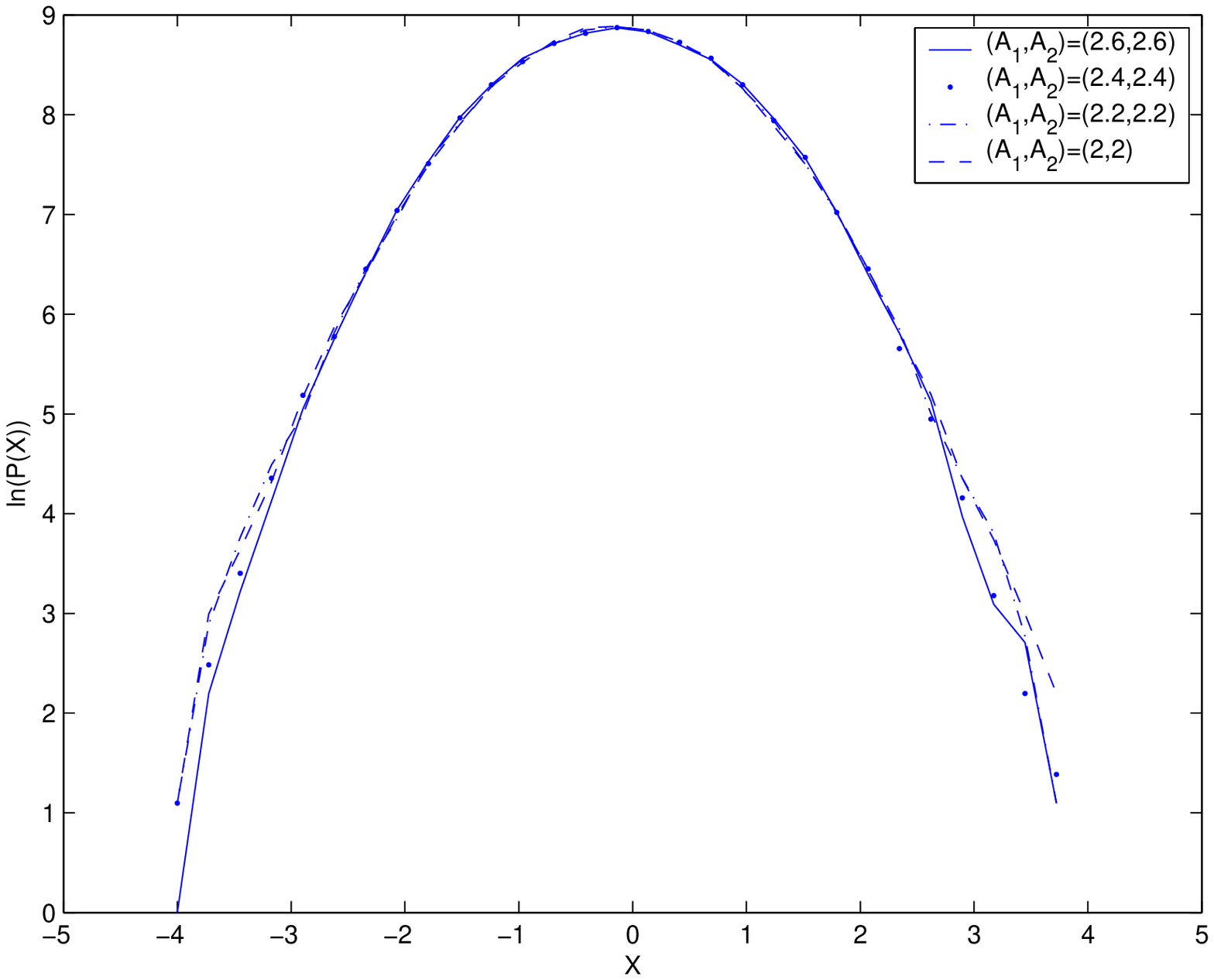}
\end{center}
\caption{pdfs for $(A_1,A_2)=(2,2)$, $(A_1,A_2)=(2.2,2.2)$, $(A_1,A_2)=(2.4,2.4)$ and $(A_1,A_2)=(2.6,2.6)$. 
Once again, note the emergence of a universal Gaussian core.}
\label{fig:fig7} 
\end{figure}

\begin{itemize}

\item Varying the diffusivity : 
Keeping $A_1,A_2$ fixed and utilizing the same checkerboard initial condition, we vary $\kappa$. 
In each case the evolution of the pdf is similar to that shown in Fig. \ref{fig:fig1}. 
The pdfs for $\kappa=2 \times 10^{-3}$ and $\kappa=5.78 \times 10^{-4}$ in the self-similar stage 
are displayed in Fig. \ref{fig:fig2}. With respect to the analytical pdf, i.e. Eq. (\ref{1l}), even 
though $\beta$ is an unknown the qualitative similarity between Fig. \ref{fig:fig2} and Fig. \ref{fig:fig0}
is evident (essentially, the dependence of $\alpha_2$ on $\kappa$ is fairly weak, when $\kappa$ changes by
an order of magnitude, as in the above simulation, $\alpha_2$ changes by a much smaller amount).
The corresponding plots of the normalized conditional dissipation are shown in Fig. \ref{fig:fig3}. Note that
for small $\kappa$, we have $G(X) \sim 1$. Also, from Fig. \ref{fig:fig3} we see that for small $\kappa$
the range of $X$ is quite small, hence in this situation the Gaussian form describes a fairly large part of the complete
pdf.
Next, in Fig. \ref{fig:fig4} we show the pdf's for a number of small diffusivities. 
Clearly, the core of $P(X)$ tends to a universal Gaussian form. 

\item Varying the strength of the flow : Fixing $\kappa=10^{-3}$ we vary $A_1,A_2$. The decay of the 
scalar variance for different flow strengths can be 
seen in Fig. \ref{fig:fig5}. Evidently, $\alpha_2 \propto $ flow
strength, therefore for a fixed $\kappa$, $\delta \propto$ 1/(flow strength). The implication being 
that the core of $P(X)$ should tend to a Gaussian function for stronger flows.
Fig. \ref{fig:fig6} shows the pdf's (in the self-similar stage) for two different flow strengths -
note the similarity to Fig. \ref{fig:fig2}. Furthermore, Fig. \ref{fig:fig7} shows the pdf's 
for a number of simulations with stronger flows. Once again, the emergence of a universal Gaussian core
is evident. Also, note the similarity to Fig. \ref{fig:fig4}. 

\end{itemize}

\section{Conclusion and Discussion}
By applying the formalism introduced by Sinai and Yakhot \cite{Sinai} to a decaying passive scalar
obeying the AD equation in a periodic domain, we obtained an expression
for the pdf of the normalized scalar field. Broadly categorized as a power-law, the core of the pdf was shown 
to be dependent 
on the physical parameters in the AD problem. Moreover, we saw the
emergence of a universal Gaussian core for the pdf in : 
(a) the limit of small diffusivity for a fixed flow strength and (b) the limit of strong flows 
for fixed $\kappa$. Combining these observations we infer the emergence of a universal Gaussian 
core in the limit of large Peclet numbers. Note that, the detailed dependence on Peclet number is not
straight forward as 
$P(X)$ is a function of both $\alpha_2$ and $\kappa$ (Eq. (\ref{1l}))  - in turn $\alpha_2$ 
depends on both on $\kappa$ and flow strength. 
Interestingly, for smaller Peclet numbers (i.e. larger $\delta$) 
the power-law pdf profile resembles a pure or stretched exponential
function. We believe that this is the reason for the mis-identification of pdf profiles in 
earlier work \cite{Ray-Chaos} (and also SP). \\

An intriguing, though poorly understood, feature of the strange eigenmode regime is 
the actual decay rate of the scalar variance - i.e. $\alpha_2$. 
Note that, in contrast to the present statistical or strange eigenmode,
when the velocity fields are {\it time periodic}, and $l_{s}(t) \sim l_{v}
\sim L$, the scalar field represents a periodic eigenfunction
of the AD operator. Hence, the structure as well as decay rate of the scalar field are better
understood. Regarding these periodic eigenfunctions, or spatially
repeating patterns, see \cite{Gol-Nat} for experimental results,
SP for a physical interpretation,
\cite{Cerbelli} for more recent work,
\cite{L-Haller} for a mathematically rigorous presentation and
\cite{Pikovsky} for an interpretation in terms of the Perron-Frobenius operator induced by the
underlying trajectory problem. Also, see \cite{Toussaint}, \cite{Gold} for similar ideas in the case 
of {\it steady}
3D and 2D flows respectively. \\ 

Before concluding we would like to put forth a plausible
connection between the eigenmode regime and homogenization theory, with the hope of shedding
some light on $\alpha_2$. 
Broadly, in the realm of homogenization theory it has been possible to show the convergence,
in a coarse grained sense, of
the AD equation to a pure diffusion equation for a variety of advecting velocity fields 
(see Section 2 of \cite{Majda-K} or \cite{Komo-review}
for recent reviews). The most common situation is when $l_v << l_s(t)$, i.e. velocity fields changing 
rapidly in space but combined with 
either steadiness or periodicity in time \cite{McLaughlin-etal},\cite{Majda-K}. The opposite
limit, along the lines of the work by Kubo \cite{Kubo}, is where the velocity fields 
have longe range spatial correlations but change rapidly in time (see section 2.4.1 of \cite{Majda-K}
or \cite{Komo-review}). \\

Indeed, it is this second limit where the effective diffusivity of the scalar field is 
given by the Taylor-Kubo formula \cite{Komo-review}. Recent work has shown that 
such a diffusive limit exists for a broader class of velocity fields, though the formula
for the effective diffusivity may not be analytically tractable \cite{Fann-Komo},\cite{Fann-2}.
Noting the long range spatial ($l_v \sim L$) and short time (randomness at each iteration) correlations of 
the velocity field required for the emergence of the self-similar eigenmode, we conjecture that
the strange eigenmode may be understood as a homogenization phenomenon. Hence, averages such as
$<\phi^n>$ obey the diffusion equation, in particular $\alpha_2$ can be interpreted as
an effective diffusivity akin to the Taylor-Kubo formula. Not only does this interpretation
lend qualitative support to the
observed dependence of $\alpha_2$ on $\kappa$ as well as the velocity field, it also implies 
the linearity of $\alpha_n$ with $n$. \\

\clearpage

\clearpage

\clearpage

\clearpage

\clearpage

\clearpage

\clearpage

\clearpage

\clearpage


\begin{thebibliography}{99}

\bibitem{Ottino} J. Ottino; {\em The Kinematics of Mixing: Stretching, Chaos, and Transport},
Cambridge University Press, 1989.

\bibitem{Ray-94} R.T. Pierrehumbert; {\em Tracer microstructure in a large-eddy dominated regime},
Chaos, Solitions and Fractals, {\bf 4}, 6, 1091, 1994.

\bibitem{Ray-Chaos} R. Pierrehumbert; {\em Lattice models of advection diffusion},
Chaos, {\bf 10}, 1, 61, 2000.

\bibitem{FH2} D. Fereday and P. Haynes; {\em Scalar decay in two-dimensional chaotic advection and
Batchelor-regime turbulence}, Submitted to Physics of Fluids, 2003.

\bibitem{me} J. Sukhatme and R.T. Pierrehumbert; {\em The Decay of Passive Scalars Under the
Action of Single Scale Smooth Velocity Fields in Bounded 2D Domains : From {\it non} self similar
pdf's to self similar eigenmodes.} Physical Review E, {\bf 66}, 056302, 2002. Erratum : {\bf 68},
019903(E), 2003.

\bibitem{Chertkov-95} M. Chertkov, G. Falkovich, I. Kolokolov and V. Lebedev ; {\em Statistics of
a passive scalar advected by a large-scale two-dimensional velocity field: Analytic solution},
Physical Review E, {\bf 51}, 5609, 1995.

\bibitem{BF-99} E. Balkovsky and A. Fouxon; {\em Universal long time properties of Lagrangian
statistics in the Batchelor regime and their application to the passive scalar problem}, Physical
Review E, {\bf 60}, 4164, 1999.

\bibitem{Falk} G. Falkovich, K. Gawedzki and M. Vergassola; {\em Particles and fields in fluid
turbulence}, Reviews of Modern Physics, {\bf 73}, 913, 2001.

\bibitem{Sinai} Ya. Sinai and V. Yakhot; {\em Limiting Probability Distributions of a Passive
Scalar in a Random Velocity Field}, Physical Review Letters, {\bf 63(18)}, 1962, 1989.

\bibitem{Ch1} E.S.C. Ching; {\em Probability Densities of Turbulent
Temperature Fluctuations}, Physical Review Letters, {\bf 70(3)}, 283, 1993.

\bibitem{Ch2} E.S.C. Ching; {\em General formula for stationary or statistically homogenous 
probability density functions}, Physical Review E, {\bf 53(6)}, 5899, 1996.

\bibitem{Pope-Ching} S. Pope and E.S.C. Ching; {\em Stationary 
probability density functions : An exact result}, Physics of Fluids, {\bf A(5)}, 1529, 1993.

\bibitem{Dopazo} C. Dopazo, L. Valino and N. Fueyo; {\em Statistical Description 
of the Turbulent Mixing of Scalar Fields}, International Journal of Modern Physics B, 
{\bf 11(25)}, 2975, 1997.

\bibitem{Pope} S. Pope; {\em Turbulent Flows},
Cambridge University Press, 2002.

\bibitem{Ching-Kraichnan} E.S.C. Ching and R. Kraichnan; {\em Exact Results For Conditional
Means of a Passive Scalar in Certain Statistically Homogenous Flows}, Journal of Statistical
Physics, {\bf 93(3-4)}, 787, 1998.

\bibitem{Vallino} L. Valino, C. Dopazo and J. Ros; {\em Quasistationary Probability Density Functions
in the Turbulent Mixing of a Scalar Field}, Physical Review Letters, 
{\bf 72(22)}, 3518, 1994.

\bibitem{Alva} M. Alvarez, F. Muzzio, S. Cerbelli, A. Androver and M. Giona; {\em Self-Similar
Spatiotemporal Structure of Intermaterial Boundaries in Chaotic Flows}, 
Physical Review Letters, {\bf 81(16)}, 3395, 1998.

\bibitem{Ant} T. Antonsen, Z. Fan, E. Ott and E. Garcia-Lopez; {\em The role of chaotic orbits in the
determination of power spectra of passive scalars}, Physics of Fluids, {\bf 8(11)}, 3096, 1996.

\bibitem{Meiss} J.D. Meiss; {\em Symplectic maps, variational principles, and transport},
Reviews of Modern Physics, {\bf 64(3)}, 795, 1992.

\bibitem{Gol-Nat} D. Rothstein, E. Henry and J. Gollub; {\em Persistent patterns in transient
chaotic mixing}, Nature, {\bf 401}, 770, 1999.

\bibitem{Cerbelli} S. Cerbelli, A. Androver and M. Giona; {\em Enhanced diffusion
regimes in bounded chaotic flows}, Physics Letters A, {\bf 312},
355, 2003.

\bibitem{L-Haller} W. Liu and G. Haller; {\em Strange Eigenmodes and Decay of Variance in the
Mixing of Diffusive Tracers}, Physica D, 188, 1, 2004.

\bibitem{Pikovsky} A. Pikovsky and O. Popovych; {\em Persistent patterns in deterministic mixing
flows}, Europhys. Lett., {\bf 61(5)}, 625, 2003.

\bibitem{Toussaint} V. Toussaint, P. Carriere, J. Scott and J-N. Gence; {\em Spectral decay
of a passive scalar in chaotic mixing}, Physics of Fluids, {\bf 12(11)},
2834, 2000.

\bibitem{Gold} I. Goldhirsch and A. Yakhot; {\em Probability-Distribution of a Passive
Scalar - A Soluble Example}, Physics of Fluids, {\bf 2(8)}, 1303, 1990.

\bibitem{Majda-K} A. Majda and P. Kramer; {\em Simplified models for turbulent diffusion:  Theory,
numerical modelling, and physical phenomena}, Physics Reports, {\bf 314 (4-5)}, 238, 1999.

\bibitem{Komo-review} T. Komorowski; {\em Transport in random media}
in {\em Proceedings of the conference on Probabilistic Problems in Atmospheric and Weather Sciences}, 
2002.

\bibitem{McLaughlin-etal} D. McLaughlin, G. Papanicolau and O. Pironneau; {\em Convection of
microstructure and related problems}, SIAM Journal of Applied Math, {\bf 45(5)}, 780, 1985.

\bibitem{Kubo} R. Kubo; {\em Stochastic Liouville Equations},
Journal of Mathematical Physics, {\bf 4(2)}, 174, 1963.

\bibitem{Fann-Komo} A. Fannjiang and T. Komorowski; {\em Turbulent Diffusion in
Markovian Flows}, The Annals of Applied Probability, {\bf 9(3)}, 591, 1999. 

\bibitem{Fann-2} A. Fannjiang and T. Komorowski; {\em Taylor-Kubo Formula for Turbulent 
Diffusion in a Non-Mixing Flow with Long-Range Correlation},
Bull. Pol. Acad. Sci., {\bf 48(3)}, 253, 2000. 

\end{thebibliography}
\end{document}